\documentclass[]{spie}

\usepackage{amsmath,amsfonts,amssymb}
\usepackage{graphicx}
\usepackage[colorlinks=true, allcolors=blue]{hyperref}
\usepackage{siunitx}
\title{Optimizing the growth conditions of Al mirrors for superconducting nanowire single-photon detectors}
\author[1,2,5,+,*]{Rasmus Flaschmann}
\author[1,2,5,+]{Christian Schmid}
\author[1,2,5]{Lucio Zugliani}
\author[1,3,5]{Stefan Strohauer}
\author[1,2,5]{Fabian Wietschorke}
\author[1,3,5]{Stefanie Grotowski}
\author[1,2,5]{Björn Jonas}
\author[3,4]{Manuel Müller}
\author[3,4,5]{Matthias Althammer}
\author[3,4,5]{Rudolf Gross}
\author[1,3,5]{Jonathan J. Finley}
\author[1,2,5]{Kai Müller}
\affil[1]{Walter Schottky Institut, Technical University of Munich, Germany}
\affil[2]{Technical University of Munich, Germany; TUM School of School of Computation, Information and Technology, Department of Electrical Engineering}
\affil[3]{Technical University of Munich, Germany; TUM School of School of Natural Sciences, Department of Physics}
\affil[4]{Walther-Meißner-Institut, Bayerische Akademie der Wissenschaften, Germany}
\affil[5]{Munich Center for Quantum Science and Technology (MCQST), Germany}

\affil[*]{rasmus.flaschmann@wsi.tum.de}

\affil[+]{these authors contributed equally to this work}

\begin{document}
\maketitle
\begin{abstract}
We investigate the growth conditions for thin ($\leq\SI{200}{\nano\meter}$) sputtered aluminum (Al) films.
These coatings are needed for various applications, e.g. for advanced manufacturing processes in the aerospace industry or for nanostructures for quantum devices.
Obtaining high-quality films, with low roughness, requires precise optimization of the deposition process.
To this end, we tune various sputtering parameters such as the deposition rate, temperature and power, which enables \SI{50}{\nano\meter} thin films with a root mean square (RMS) roughness of less than \SI{1}{\nano\meter} and high reflectivity.
Finally, we confirm the high-quality of the deposited films by realizing superconducting single-photon detectors integrated into multi-layer heterostructures consisting of an aluminum mirror and a silicon dioxide dielectric spacer. We achieve an improvement in detection efficiency at \SI{780}{\nano\meter} from \SI{40}{\percent} to \SI{70}{\percent} by this integration approach.
\end{abstract}

\section*{Introduction}

In recent years, photon-based quantum technologies have made great progress in various fields \cite{Kim08}, such as deep space optical communication (DSOC) using photons \cite{Cal19, Iva20}, quantum computation \cite{Sin16}, on-chip photonic circuits \cite{Wan20, Ter22}, or quantum key distribution (QKD) \cite{BB84, Shi14}.
Such applications require special components such as single-photon emitters (e.g., NV centers in diamond \cite{Kur00}, 2D materials \cite{Tra16}, or quantum dots \cite{Sen17}), spin-photon interfaces \cite{Ata18}, and detectors.
For the latter, superconducting (nanowire) single-photon detectors (SNSPDs) \cite{Gol01, Rei13, Rei13a, Rei15, Red16, Fla22a, Fla23a} have prevailed over other potential candidates such as transition edge sensors (TES) \cite{Ull15} or single-photon avalanche diodes (SPADs) \cite{Nix32}.
In particular, SNSPDs outperform the other systems by their excellent timing resolution in the range of a few \si{\pico\second} \cite{Kor20} combined with their high system detection efficiency in the visible to near-infrared \cite{Red20} when integrated into a resonator structure.
Furthermore, first detectors have also been tested in the ultra-violet range \cite{Wol17}.
However, since the thickness of these detectors is typically in the range of a few \si{\nano\meter} \cite{Wol20}, the structural properties at the interface between the detector and substrate are very demanding.
Moreover, their detection efficiency strongly depends on the resonator in which they are integrated.
For this purpose, multi-layer structures (e.g. distributed Bragg reflectors (DBR) \cite{Zha17} or broadband resonators \cite{Red20}) can be used to increase absorption and thus system efficiency.
To ensure a clean surface, the materials used to produce a single-sided resonator (here: aluminum mirror and silicon dioxide spacer) underneath the detector are ideally deposited in the same ultra-high vacuum (UHV) chamber as the superconductor and exhibit a combined RMS roughness of less than \SI{0.25}{\nano\meter}.
Beyond applications in resonators for SNSPDs, flat aluminum coatings are used for many purposes such as echelle gratings \cite{Liz15}, advanced manufacturing technologies including the aerospace industry \cite{Xie21} or applications in integrated circuits \cite{Rob86}.
Hence, improving the fabrication process of aluminum thin films has the potential to lead to advancements in various fields.
%SNSPDs have proven outstanding performance metrics for measuring single-photons.

%\clearpage
\section*{Results}

Due to their high sensitivity over a wide wavelength range, broadband resonators are of particular interest.
For the fabrication of such broadband resonators a suitable mirror material has to be found that ideally can be deposited in the same UHV chamber as the superconductor while at the same time exhibiting a surface roughness of less than \SI{1}{\nano\meter}.
To determine suitable mirror materials, we consider gold and aluminum as two potential candidates and compare them to silicon as a standard semiconductor substrate.

\subsection*{Influence of the deposition rate}

\begin{figure}[ht]
\centering
\includegraphics[height=6cm]{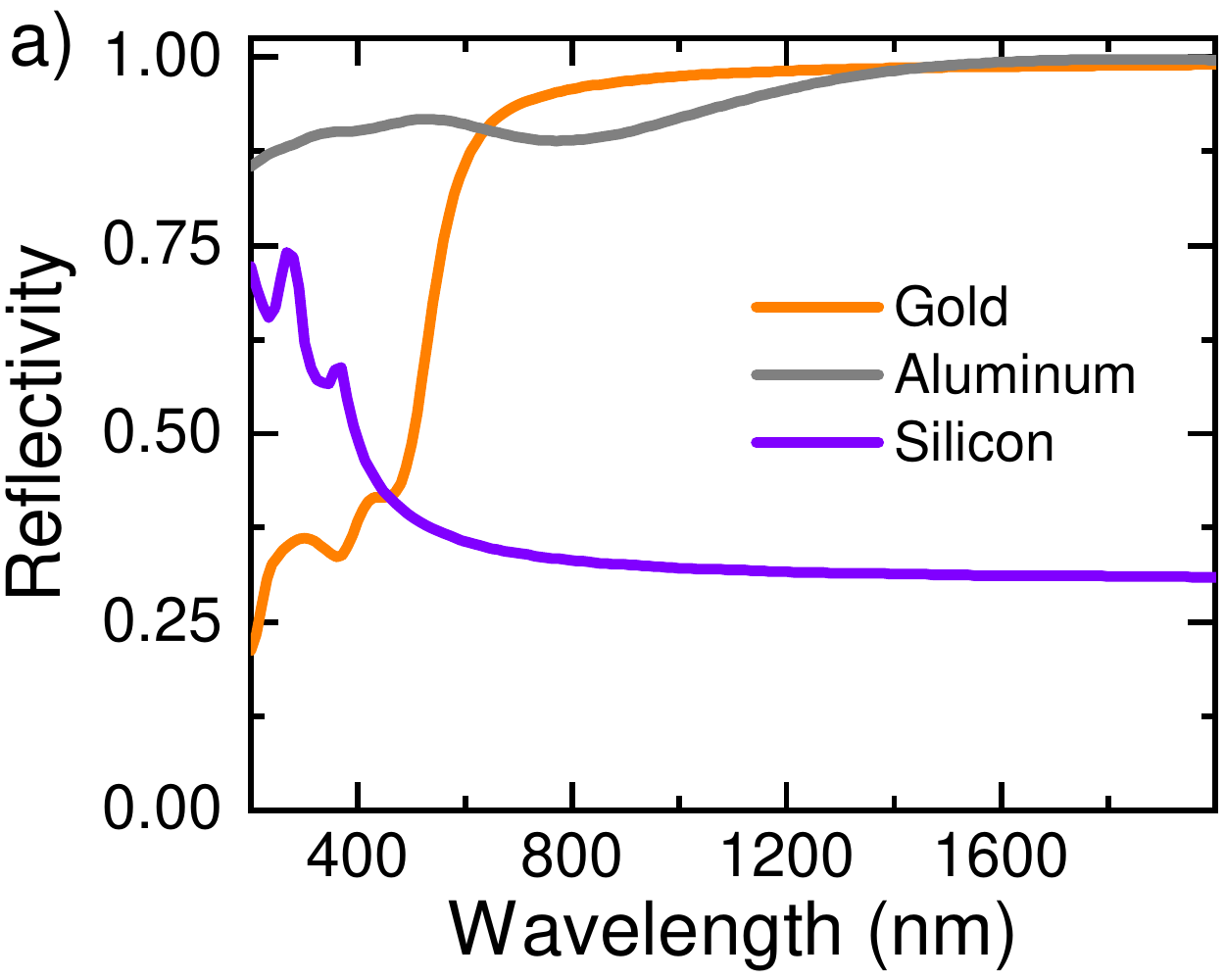}
\includegraphics[height=6cm]{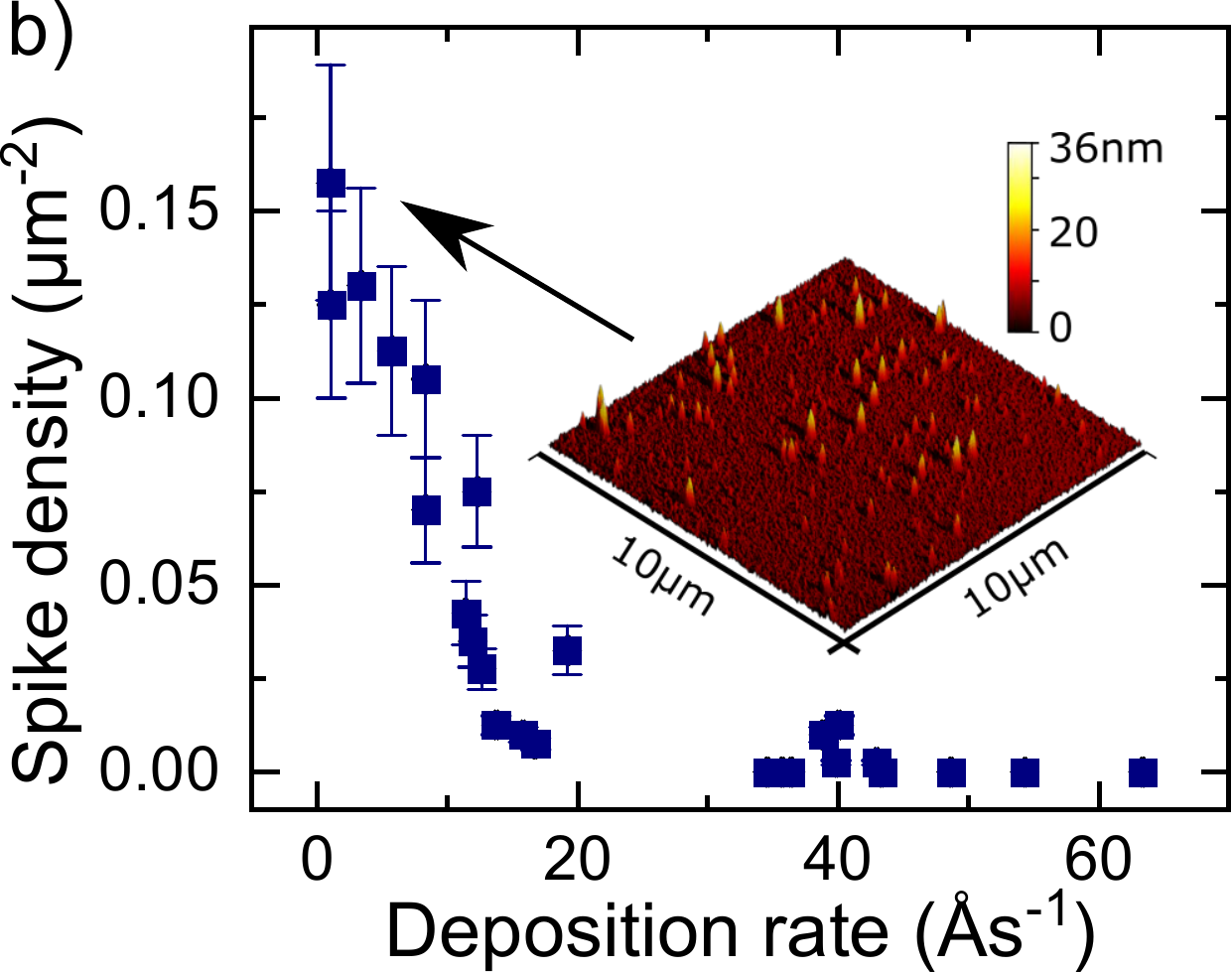}
\caption{Reflectivity and spike density of potential substrate materials. (a) Measurement of the reflectivity of Au, Al, and Si as a function of the wavelength. While aluminum has the highest reflectivity between \SI{200}{\nano\meter} and \SI{600}{\nano\meter}, it is surpassed by gold between \SI{600}{\nano\meter} and \SI{1200}{\nano\meter}. At even higher wavelengths, gold and aluminum behave almost identically with a reflectivity over \SI{99}{\percent}. Only the silicon wafer exhibits a significantly lower reflectivity at wavelengths above \SI{400}{\nano\meter}. (b) Spike density as a function of the deposition rate for sputtered aluminum films, as determined from AFM surface topography scans. The inset shows an AFM image for a sample with a high spike density sputtered at a rate of \SI{1}{\angstrom\per\second}. The error bars spike density represent the standard deviation from multiple AFM scans on a sample.}
\label{fig:1}
\end{figure}

Fig. \ref{fig:1}(a) shows the measured reflectance spectrum of these three materials as a function of the operating wavelength.
The results were obtained using a thin film reflectometer (Filmetrics F20) (see Methods section for more details).
The recorded data reveal that silicon is not a suitable material for achieving high detection efficiencies due to its low reflectivity above \SI{400}{\nano\meter} ranging between 0.30 and 0.40.
We therefore focus on the two remaining materials.
While aluminum has the highest reflectivity between \SI{200}{\nano\meter} and \SI{600}{\nano\meter}, it is surpassed by gold between \SI{600}{\nano\meter} and \SI{1200}{\nano\meter}.
At even higher wavelengths, gold and aluminum behave almost identically with a reflectivity over \SI{99}{\percent}.
Hence, a sufficient reflectivity can be achieved by either of the two possible materials.
However, sputtering gold \cite{Xin10} can cause cross contamination and therefore influences other sputtering processes performed in the same chamber.
Consequently, we focus on developing a sputter deposition process for high-quality aluminum coatings that combines high reflectivity with low surface roughness. 
Throughout this work a \SI{525}{\micro\meter} thick Si-wafer with a \SI{127}{\nano\meter} thick thermally grown SiO$_2$ layer on top was used as a substrate material.
For these substrates, the RMS surface roughness was determined to be \SI{0.25}{\nano\meter}, which is close to the resolution limit of the atomic force microscope (AFM).

We start the film optimization by investigating the large peaks present in the deposited material, hereafter referred to as spikes.
The inset of Fig. \ref{fig:1}(b) shows a typical atomic force microscope (AFM) image for a \SI{50}{\nano\meter} thick Al film deposited by DC magnetron sputtering with a rate of \SI{1}{\angstrom\per\second}.
A detailed description of the acquisition of the AFM images is given in the Methods section.
The scan has a size of \SI{10}{\micro\meter} $\times$ \SI{10}{\micro\meter} corresponding to the typical area covered by a superconducting detector.
Here, individual spikes can be clearly identified with a maximum height of \SI{36}{\nano\meter}.
The spikes are defined as areas that are more than \SI{10}{\nano\meter} above the mean height level of the surface.
The spikes are likely caused by residuals of oxygen or water in the atmosphere of the sputtering chamber during deposition \cite{Ver86, Cha91} or cross-contamination.
Fig. \ref{fig:1}(b) shows the measured spike density as a function of deposition rate for various aluminum films.
Here, the rate was tuned by various parameters such as the applied power, the distance between the target and substrate holder or the argon sputtering pressure.
The data reveal that the spike density decreases with an increased deposition rate up to around \SI{50}{\angstrom\per\second}.
At even higher deposition rates, the spike density is no longer observable.
This indicates that the density of spikes depends on the time required for the formation of a monolayer (monolayer formation time).
It describes the time required to cover the sample surface by residual substances present in the deposition chamber \cite{Qui21}.
At higher deposition rates, the amount of impurities incorporated into the layer decreases, resulting in lower stress and fewer spikes.
In addition, we found that cleaning the sputtering target by pre-sputtering prior to deposition is an indispensable step.
To investigate this, we produced four mirrors directly one after the other without any cleaning process (sputtering time: \SI{10}{\second}) and measured the corresponding spike density.
It decreased from \SI{0.65 \pm 0.02}{\per\micro\meter\squared} (S1) over \SI{0.0525 \pm 0.02}{\per\micro\meter\squared} (S2) and \SI{0.025 \pm 0.02}{\per\micro\meter\squared} (S3) down to \SI{0.02 \pm 0.02}{\per\micro\meter\squared} (S4).
With increasing sputtering time, contamination is removed from the system and the film quality is improved.
This emphasizes the need for a pre-sputtering process prior to sputter deposition, as this in turn leads to less stress and fewer spikes.
%%In particular, we found that spikes increase significantly if the Al target is also used for other materials such as AlN deposited via reactive sputtering with a nitrogen-argon gas mixture.
%%Nitration of the target surface leads to contamination of the Al target, increasing the probability of spike formation.
%%Consequently, prior to the deposition of the aluminum thin film we clean the Al target by sputtering with closed shutter for at least \SI{10}{\min}.
%%However, it must be considered that heat is generated during this process, which increases the temperature in the sputtering chamber and thereby degrades the surface roughness.
%%Hence, it is necessary to wait at least another \SI{10}{\min} to let the target cool down again before the Al thin film can be applied.
%\clearpage

\begin{table}[ht]
\centering
\begin{tabular}{|l|l|l|l|}
\hline
Distance to target (\si{\milli\meter}) & RMS roughness (\si{\nano\meter}) & Grain size ($10^{-3}$ \si{\micro\meter\squared}) & Deposition rate (\si{\angstrom\per\second}) \\
\hline
 40 &  $1.34 \pm 0.25$ & $3.10 \pm \SI{5}{\percent}$ & $38.9 \pm 0.5$ \\ \hline
 80 &  $1.76 \pm 0.25$ & $3.71 \pm \SI{5}{\percent}$ & $12.6 \pm 0.5$\\ \hline
120 &  $1.82 \pm 0.25$ & $4.23 \pm \SI{5}{\percent}$ &  $5.8 \pm 0.5$\\ \hline
\end{tabular}
\caption{\label{tab:1} Influence of the distance between the target and the substrate holder. With decreasing distance, a decreasing surface roughness and grain size is obtained. At the same time, the deposition rate increases significantly as well.
The uncertainty of the rate stems from the ratio of the measured deposition rate given by the deposition system and the film thickness determined via X-ray reflectometry (XRR).}
\end{table}

In addition, we investigated the influence of the distance between the target and substrate holder on the deposition process for \SI{50}{\nano\meter} thick films as shown in Tab. \ref{tab:1}.
In particular, we observed an increased deposition rate for decreasing distance.
This indicates that a low distance such as \SI{40}{\milli\meter} (system dependent minimum) is preferred as it avoids the formation of spikes (cf. Fig. \ref{fig:1}(b)).
We note that both the RMS roughness and grain size decreased with decreasing distance as well.
To conclude, it is necessary to clean the target from contamination prior to deposition and it is advantageous to use a high deposition rate (for instance by reducing the distance between the substrate holder and target) to achieve high-purity films without spikes.

\subsection*{Impact of the deposition temperature}
\begin{figure}[ht]
\centering
\includegraphics[height=6cm]{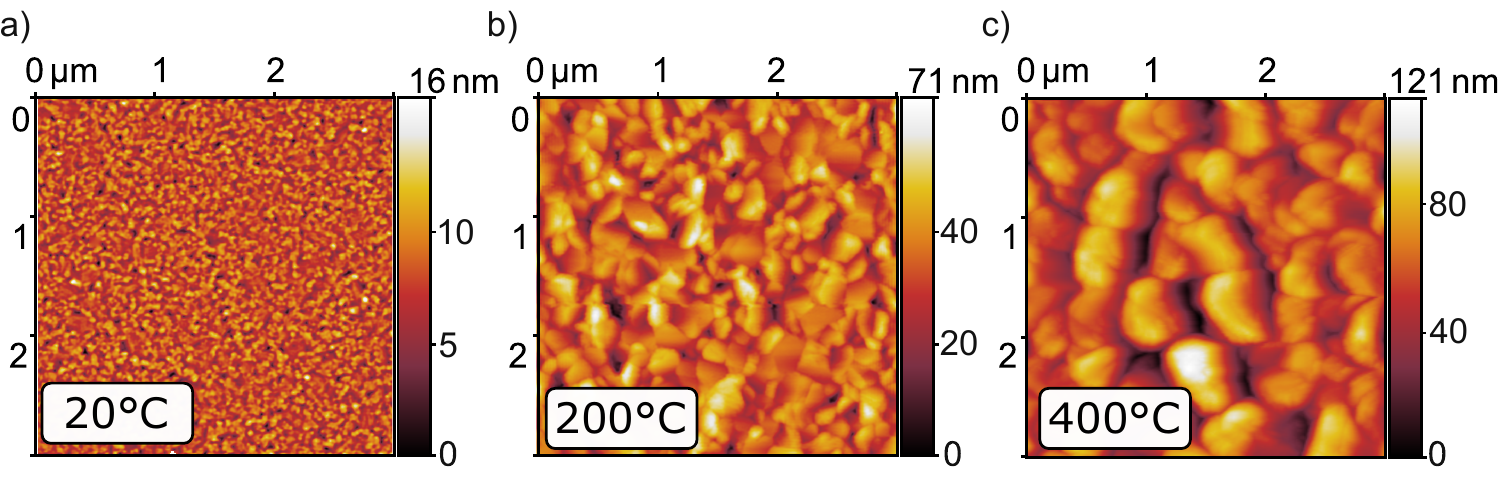}
\caption{Impact of deposition temperature on the structural properties of aluminum thin films determined by AFM measurements.
(a) While the films deposited at room temperature have many small grains, the grain size increases with increasing deposition temperature leading to larger features, as shown in the images (b) at \SI{200}{\celsius} and (c) at \SI{400}{\celsius}. 
In addition, the maximum height variation also increases from \SI{16}{\nano\meter} for films deposited at room temperature, over (b) \SI{71}{\nano\meter} at \SI{200}{\celsius} up to (c) \SI{121}{\nano\meter} at \SI{400}{\celsius}.
Hence, a higher deposition temperature leads to larger grains and a higher surface roughness indicating that a room temperature deposition process should be used.}
\label{fig:2}
\end{figure}

As a next parameter, we investigated the impact of the substrate temperature during the deposition of \SI{50}{\nano\meter} aluminum films in the range from \SIrange{20}{400}{\celsius}.
The corresponding AFM images covering a size of \SI{3}{\micro\meter} $\times$ \SI{3}{\micro\meter} are shown in Fig. \ref{fig:2}(a) - Fig. \ref{fig:2}(c).

Fig. \ref{fig:2}(a) shows a homogeneous film with a maximum height variation of about \SI{16}{\nano\meter} deposited at room temperature using non-optimized fabrication parameters with a power of \SI{200}{\watt}, a distance of \SI{80}{\milli\meter} at a pressure of \SI{5E-3}{\milli\bar}.
However, for an increased deposition temperature in Fig. \ref{fig:2}(b) and Fig. \ref{fig:2}(c), the height variation increases even further up to \SI{71}{\nano\meter} and \SI{121}{\nano\meter}, respectively.
Also, the general surface morphology of the presented aluminum films changed significantly as presented in Tab. \ref{tab:2}.

\begin{table}[ht]
\centering
\begin{tabular}{|l|l|l|}
\hline
Temperature (\si{\celsius}) & RMS roughness (\si{\nano\meter}) & Grain size ($10^{-3}$ \si{\micro\meter\squared}) \\
\hline
 20 &  $1.78 \pm 0.25$ &   $4.5 \pm \SI{5}{\percent}$ \\ \hline
200 &  $9.22 \pm 0.25$ &  $62.7 \pm \SI{5}{\percent}$ \\ \hline
400 & $17.21 \pm 0.25$ & $223.5 \pm \SI{5}{\percent}$ \\ \hline
\end{tabular}
\caption{\label{tab:2} Obtained roughness and grain size of a \SI{50}{\nano\meter} thick Al film derived from the AFM images shown in Fig. \ref{fig:2}(a) - Fig. \ref{fig:2}(c). The RMS roughness (from \SIrange{1.8}{17.2}{\nano\meter}) and grain size (from \SI{4.5E-3}{\micro\meter\squared} to \SI{223.5E-3}{\micro\meter\squared}) increase significantly with increasing deposition temperature (from \SIrange{20}{400}{\celsius}).}
\end{table} %SI{4.54 E-3}{\micro\meter\squared} to \SI{223.49 E-3}{\micro\meter\squared})

We observe that the sputtered films exhibit larger grains (average values range from \SI{4.5E-3}{\micro\meter\squared} to \SI{223.5E-3}{\micro\meter\squared}) and become rougher (from \SIrange{1.8}{17.2}{\nano\meter}) for an increased deposition temperature (from \SIrange{20}{400}{\celsius}).
We deduce that the grain size and RMS roughness are correlated.
In particular, by reducing the grain size the surface roughness is improved.
Moreover, the findings indicate that only films sputtered at ambient temperature are suitable for the fabrication of high quality thin films.

\subsection*{Impact of Ar sputtering pressure and applied power}
To investigate the impact of the deposition rate in more detail we subsequently analyzed the influence of the argon pressure during the deposition of \SI{75}{\nano\meter} thick aluminum films.

\begin{figure}[ht]
\centering
\includegraphics[height=6cm]{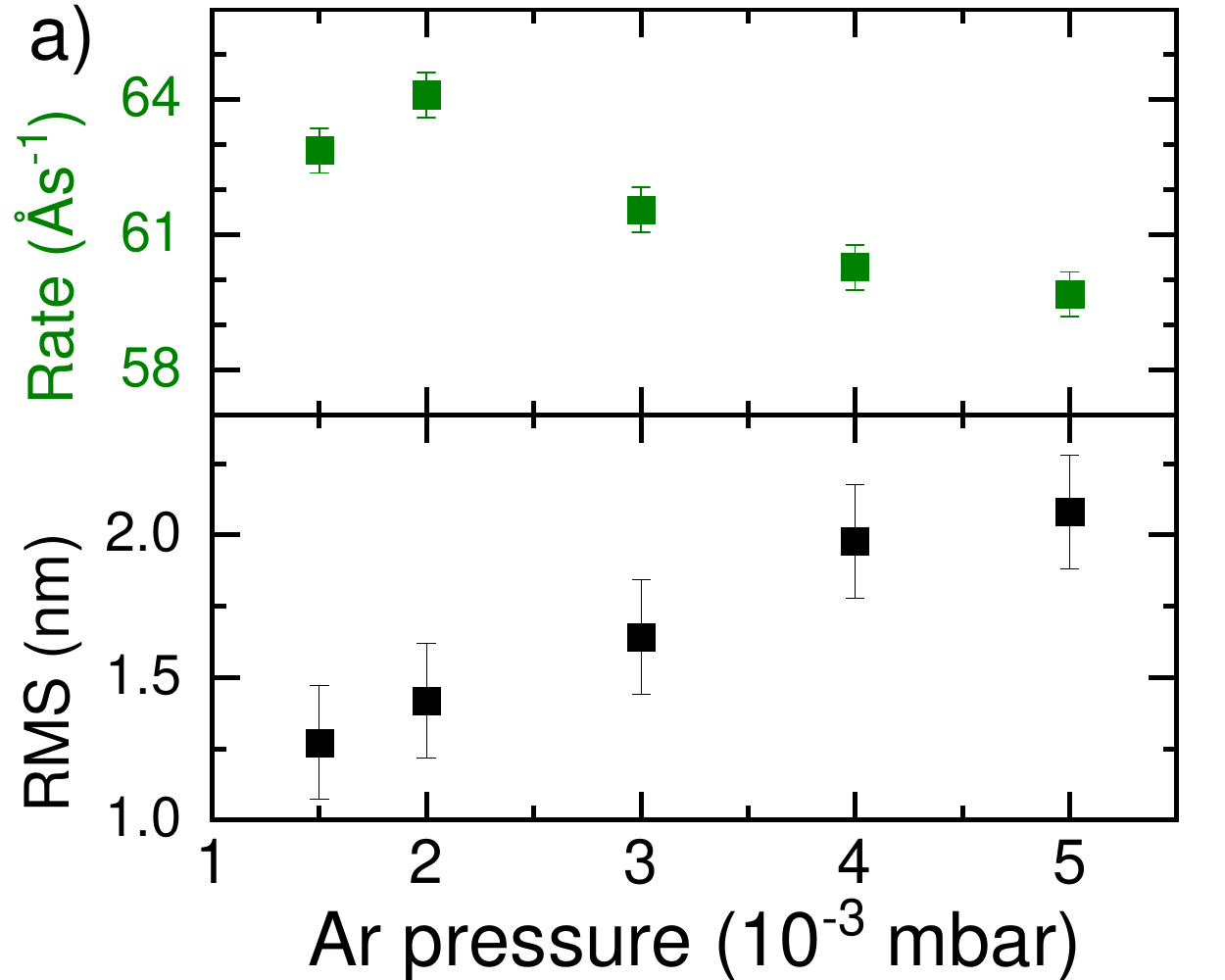}
\includegraphics[height=6cm]{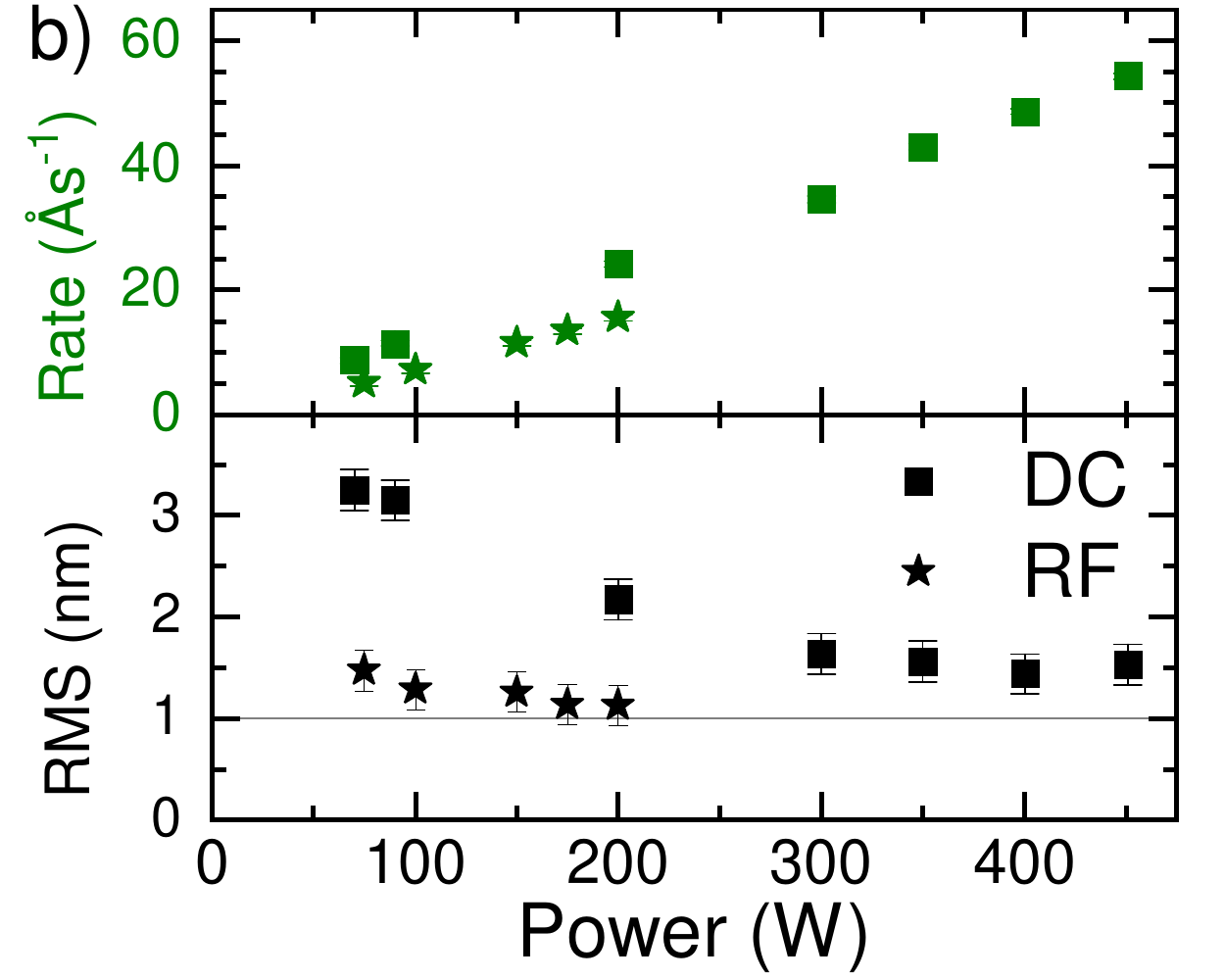}
\caption{Influence of the deposition rate on the thin film quality. (a) Deposition rate and RMS roughness as a function of argon sputtering gas pressure at a constant power of \SI{450}{Watt}. Increased rate and improved roughness are observed at reduced pressure. (b) Comparison of deposition rate and surface roughness as a function of applied power for both DC and RF sputtering. For DC sputtering the roughness improves with increasing power at a constant sputtering pressure of \SI{3E-3}{\milli\bar}. However, RF sputtering allows for even better roughness values at comparatively lower powers.}
\label{fig:3}
\end{figure}

Fig. \ref{fig:3}(a) shows the corresponding RMS roughness and deposition rate as a function of the argon pressure.
With an increasing Ar pressure (from \SI{1.5E-3}{\milli\bar} to \SI{5.0E-3}{\milli\bar}) the deposition rate decreases (\SIrange{64.2}{59.6}{\angstrom\per\second}) likely due to an increased collision probability between the aluminum and argon atoms and a drop in DC bias voltage.
Simultaneously, an increased roughness (from \SIrange{1.3}{2.2}{\nano\meter}) can be observed.
Note that the applied DC power of \SI{425}{\watt} has been selected to avoid the formation of spikes (cf. Fig. \ref{fig:1}(b)).
Also here, an increased deposition rate resulted in a lower surface roughness and therefore higher quality films.
It should be noted, however, that at even lower pressures the plasma is no longer stable.
Hence, an argon pressure around \SI{2E-3}{\milli\bar} is the best parameter for our system.
We then investigated the influence of the applied power for both DC and RF magnetron sputtering configuration.
%\textcolor{red}{can also be cut in my opinion}
%\textcolor{red}{Note that for DC magnetron sputtering, the argon ions are directly accelerated towards the target by a constant electric field allowing them to extract the aluminum.
%In contrast, for RF sputtering the argon ions and electrons are accelerated alternately in both directions. 
%Above a frequency of \SI{50}{\kilo\hertz}, the ions can no longer follow the alternating field because their charge-to-mass ratio is too low.
%The electrons, on the other hand, oscillate in the plasma and collide more often with argon atoms.
%The positive ions move toward the target through a superimposed negative offset voltage, where they detach aluminum atoms by collisions.}
%%Subsequent deposition is performed as in DC sputtering. 
Note that one of the main advantages of RF sputtering is that it produces significantly less heat.
Fig. \ref{fig:3}(b) shows the roughness (bottom) and deposition rate (top) as a function of the applied DC and RF power for an argon pressure of \SI{5E-3}{\milli\bar} and both power configurations.
Here, the data points obtained for DC sputtering are shown as squares and those for RF sputtering as stars.
Starting with DC sputtering, we observe that an increased rate is accompanied by a decreased roughness down to \SI{1.5}{\nano\meter} at \SI{425}{\watt} (system dependent maximum).
However, by using RF sputtering, we were able to improve the roughness even further, from \SI{1.47}{\nano\meter} at \SI{75}{\watt} to \SI{1.12}{\nano\meter} at \SI{200}{\watt} (system dependent maximum for RF sputtering).
We attribute the observed improvement in surface roughness to the reduced heating of the substrate by the RF sputtering process.
This is in good agreement with the results from Fig. \ref{fig:2}, which showed a direct correlation between the deposition temperature and measured film roughness.
Therefore, to achieve the best possible surface roughness, it is advantageous to use an argon pressure in the range of \SI{2E-3}{\milli\bar} in combination with the RF magnetron sputtering configuration at an applied power of \SI{200}{\watt}.

\subsection*{Impact of the aluminum layer thickness on the thin film quality}
\begin{figure}[ht]
\centering
\includegraphics[height=6cm]{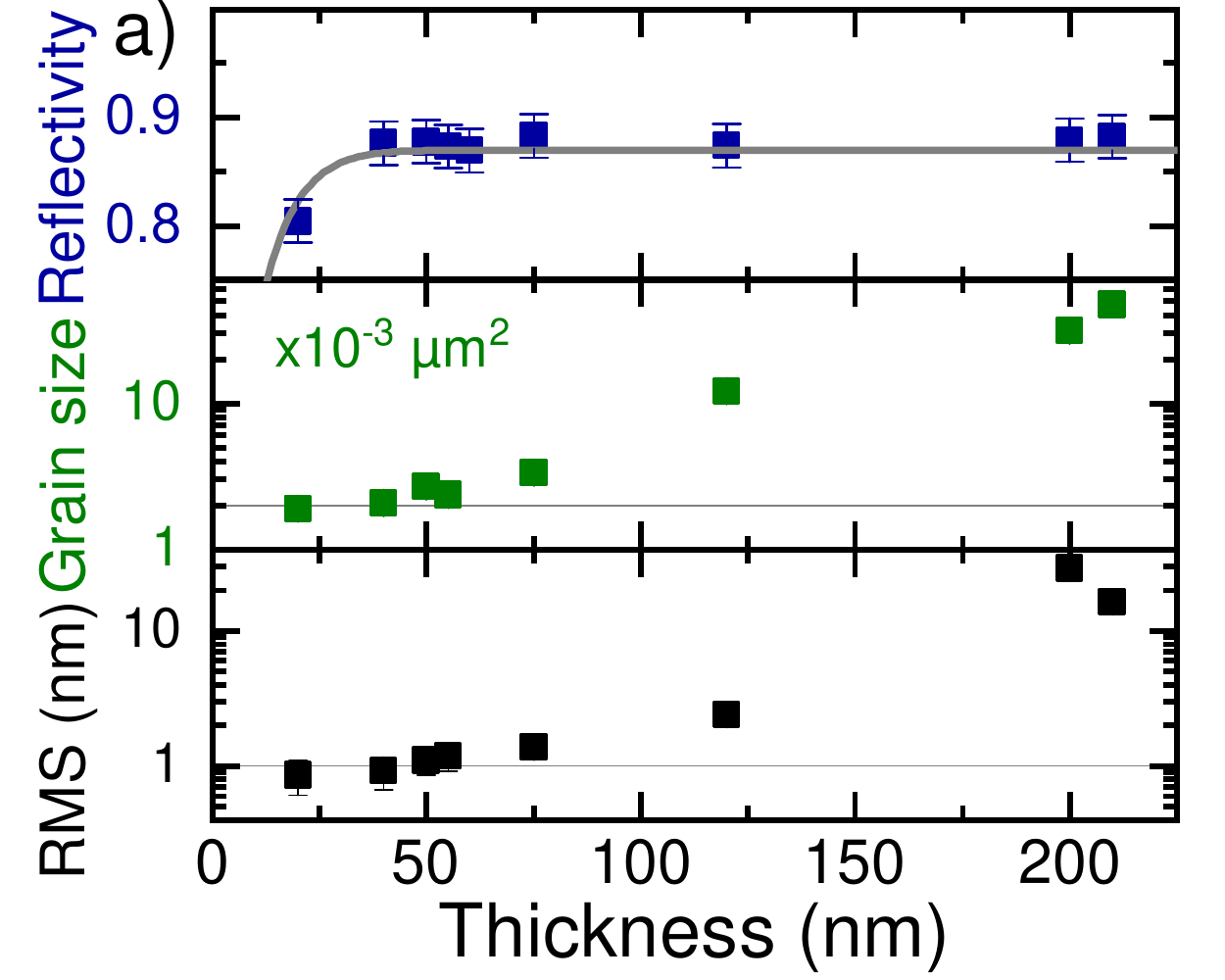}
\includegraphics[height=6cm]{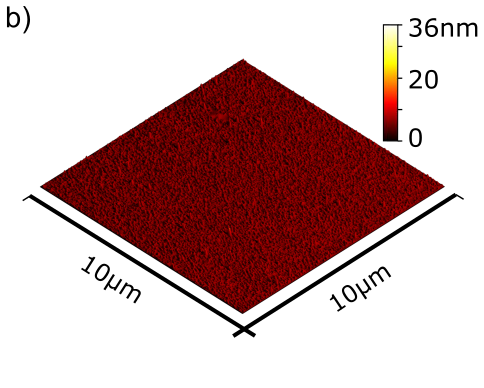}
\caption{Influence of film thickness on the surface quality for coatings produced by the optimized manufacturing process. (a) Dependence of surface roughness, grain size and reflectivity on the aluminum thickness, plotted on a semi-logarithmic scales.
With increasing film thickness, both roughness and grain size increase, while the reflectivity saturates with increasing thickness.
(b) AFM image of a sample prepared by the optimized fabrication method shows a low surface roughness and no spikes.}
\label{fig:4}
\end{figure}

Finally, we produced and analyzed aluminum films with varying layer thicknesses using the previously optimized fabrication parameters.
The corresponding results in Fig. \ref{fig:4}(a) show the measured reflectivity at a wavelength of \SI{780}{\nano\meter}, grain size and RMS roughness as a function of the sputtered aluminum thickness.
For a thickness above \SI{50}{\nano\meter}, the obtained data reveals a saturating reflectivity at \SI{85}{\percent}.
Moreover, with an increasing film thickness, both roughness and grain size, plotted on a semi-logarithmic scale, increase significantly.
This indicates that small film thicknesses are preferable for ultra-flat coatings.
In particular, for film thicknesses below \SI{50}{\nano\meter} we obtain a roughness of less than \SI{1}{\nano\meter} as well as a grain size of less than \SI{2E-3}{\micro\meter\squared} indicating a high film homogeneity.
Fig. \ref{fig:4}(b) shows a corresponding AFM scan with a low surface roughness and no spikes.
If we now combine the previously determined results, we conclude that an aluminum layer thickness of \SI{50}{\nano\meter} should be used.
It allows us to fabricate smooth thin films with a grain size of less than \SI{2E-3}{\micro\meter\squared} accompanied by a low surface roughness ($<\SI{1}{\nano\meter}$ RMS) and saturating reflectivity.

\subsection*{Superconducting detectors on aluminum mirrors}
To confirm the high quality of the optimized aluminum mirrors, we benchmark superconducting nanowire single-photon detectors (SNSPDs) fabricated on an aluminum-silica one-sided resonator.
The multi-layer heterostructure consists of an aluminum mirror buried below a silicon dioxide spacer on top of which the NbTiN detector is fabricated.
These detectors consist of a meandering nanowire with a typical wire width around \SI{100}{\nano\meter}, a thickness below \SI{10}{\nano\meter} and are cooled down well below the superconductors critical temperature.
When now a current is applied near the so-called critical current, the energy of a single photon is sufficient to destroy the superconducting state and a normal conducting barrier is formed.
The voltage pulse generated by this process can then be amplified and measured \cite{Gol01, Nat12}.

\begin{figure}[ht]
\centering
\includegraphics[height=6cm]{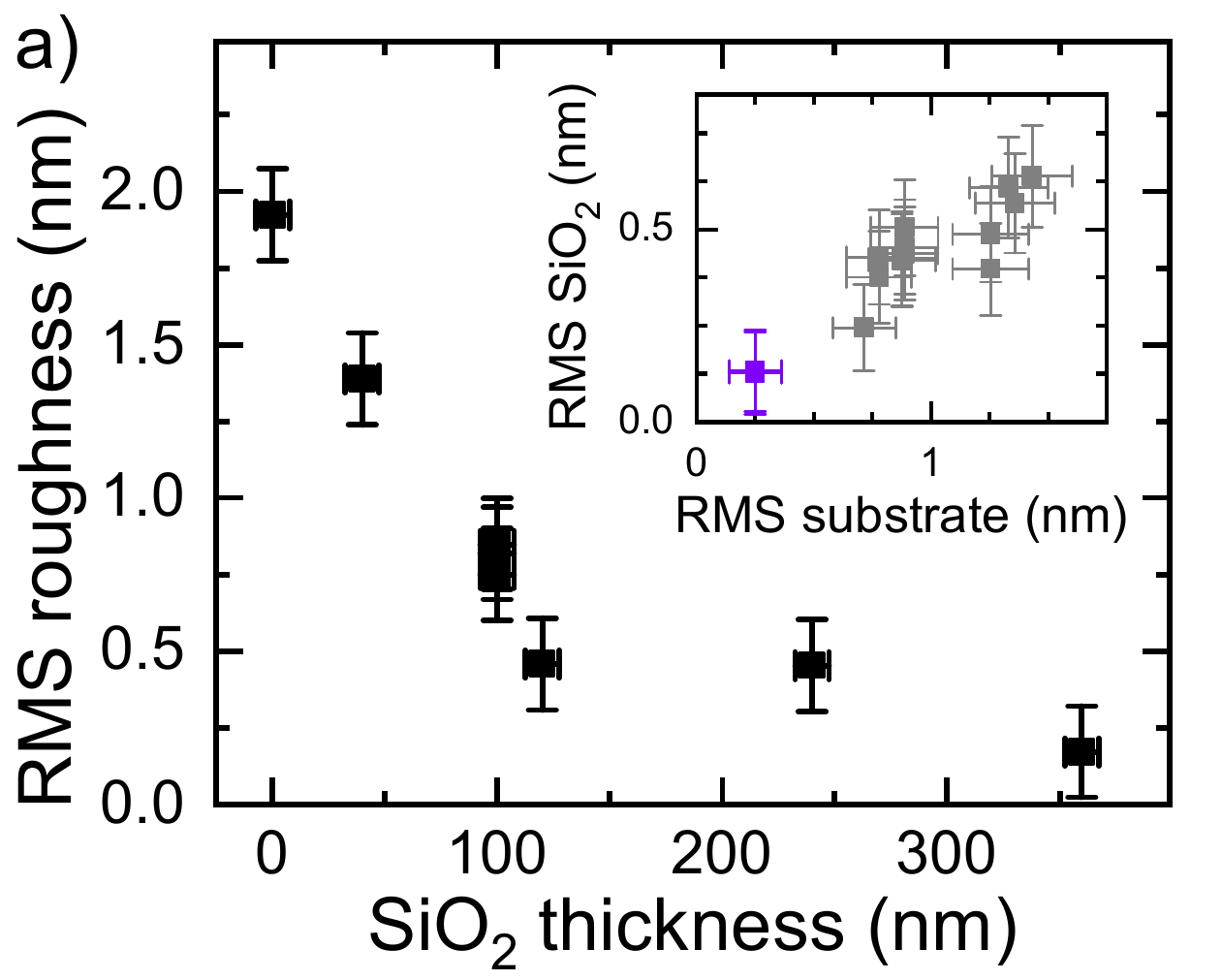}
\includegraphics[height=6cm]{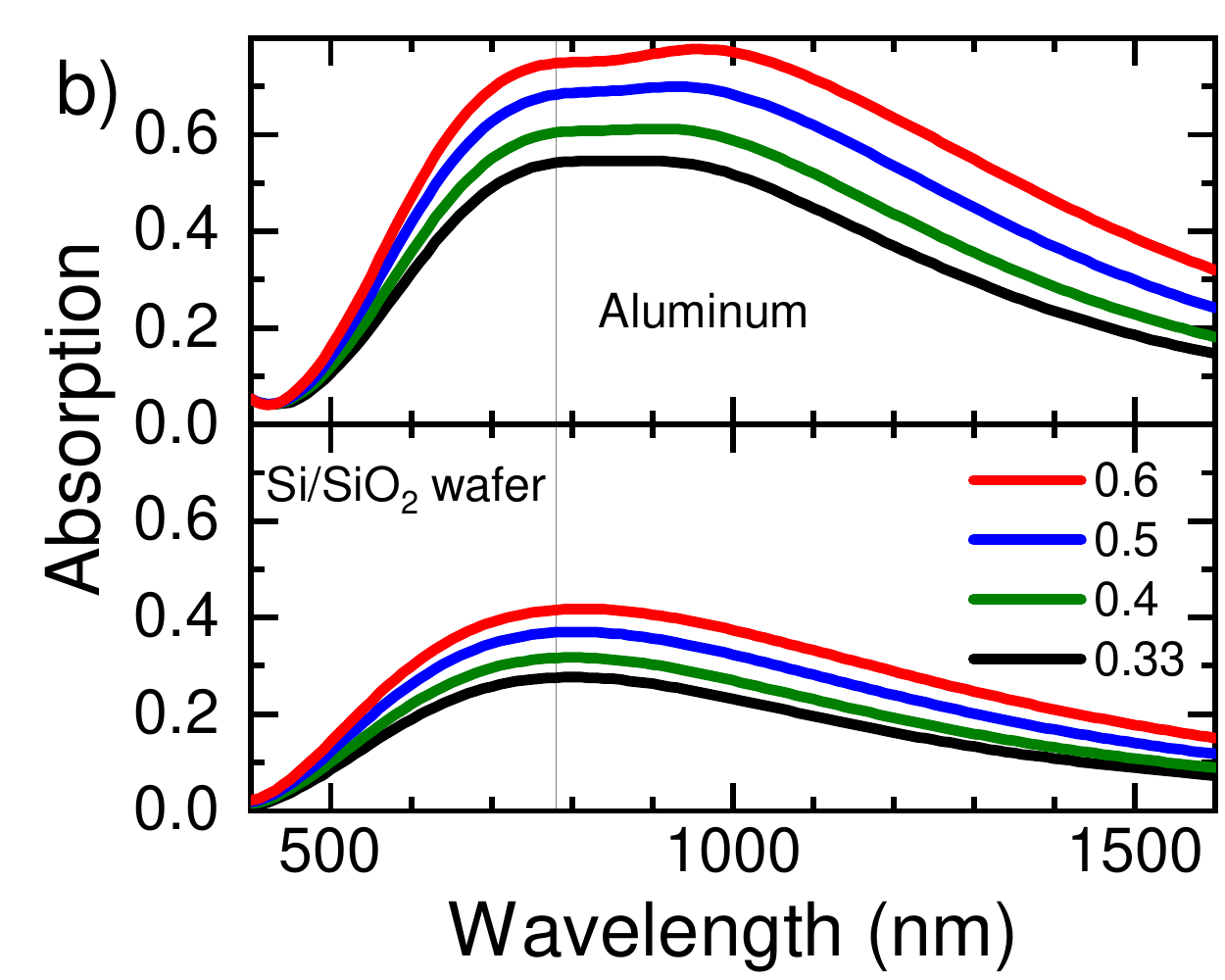}
\caption{Integration of superconducting single-photons detectors into a single-sided resonator consisting of an aluminum mirror and a SiO$_2$ spacer. (a) The surface roughness can be improved by introducing a SiO$_2$ top layer. For a film thickness of \SI{120}{\nano\meter}, the roughness decreased from \SI{1.92}{\nano\meter} down to \SI{0.46}{\nano\meter}. The inset shows the improvement of films with different roughness before and after the deposition of \SI{120}{\nano\meter} SiO$_2$. (b) Simulated absorption of detectors with varied fill factor from 0.33 (black) to 0.6 (red) on the semi-sided aluminum resonator with a dielectric, \SI{120}{\nano\meter} thick SiO$_2$ spacer on top of a Si/SiO$_2$ substrate (top panel). Simulated detector absorbance on bare silicon wafer with \SI{127}{\nano\meter} thermally grown SiO$_2$ only (bottom panel). The reference line marks the operating wavelength of \SI{780}{\nano\meter} used in this work.
The data reveal an increased absorption probability for the semi-sided aluminum resonator independent of the wavelength.
The top panel reveals a broad spectral range of high absorption in the range from \SIrange{750}{1000}{\nano\meter} for a SiO$_2$ thickness optimized for operation at \SI{780}{\nano\meter}.}
\label{fig:5}
\end{figure}

The roughness of the fabricated multi-layer structure consisting of the aluminum film and a silicon dioxide layer with a thickness in the range from \SIrange{0}{360}{\nano\meter} is shown in Fig. \ref{fig:5}(a).
%Note that for
For this study we did not use the fully-optimized parameters in order to analyze the impact of the SiO$_2$ thickness in more detail.
For an initial roughness of \SI{1.92}{\nano\meter} we observe a strong decrease down to \SI{0.46}{\nano\meter} for a silica layer thickness of approximately \SI{120}{\nano\meter}.
For even thicker silica films only small improvements can be seen.
Note that the initial roughness of the wafer itself is around \SI{0.25}{\nano\meter}.
%Here, a similar parameter set compared to aluminum has been used for the SiO$_2$ deposition.
The inset of Fig. \ref{fig:5}(a) depicts the improvement of films with different roughness before and after the deposition of \SI{120}{\nano\meter} SiO$_2$.
Especially the aluminum mirrors (grey) show that the roughness improves by a factor of two even for films that already had an initial roughness below \SI{1}{\nano\meter}.
Hence, by combining the optimized deposition processes for aluminum and silicon dioxide we are able to fabricate a broadband resonator with an RMS roughness comparable to a commercially available silicon dioxide-on-silicon wafer around \SI{0.25}{\nano\meter}.
Fig. \ref{fig:5}(b) presents the simulated absorption as a function of the operating wavelength.
In the bottom panel, a silicon wafer with a \SI{127}{\nano\meter} thick thermally grown SiO$_2$ layer was used.
Here, the absorption increases with the fill factor and shows a fill factor-dependent maximum at a wavelength of \SI{780}{\nano\meter} corresponding to the design wavelength of the optical stack.
The data reveal a maximum absorption of \SI{27.6}{\percent} for a fill factor of 0.33 and an absorption of \SI{41.6}{\percent} for a fill factor of 0.6.
In the top panel, the results for a semi-sided resonator of \SI{120}{\nano\meter} SiO$_2$ on top of \SI{50}{\nano\meter} Al are shown.
The absorption probability increases with an increasing fill factor as well and is almost doubled compared to the Si/SiO$_2$ wafer.
Here, absorption values of \SI{54.2}{\percent} (fill factor 0.33) and \SI{74.8}{\percent} (fill factor 0.6) were determined.
The SiO$_2$ thickness was chosen according to the simulation results and formed a $\lambda/4$ resonator for a refractive index $n = 1.46$ of SiO$_2$ \cite{Bor12}.
Furthermore, note the broad spectral range of high absorption from \SIrange{750}{1000}{\nano\meter}.
To conclude, the obtained results show that the aluminum resonator improves the absorption probability by forming a broad optical resonator.

\begin{figure}[ht]
\centering
\includegraphics[height=6cm]{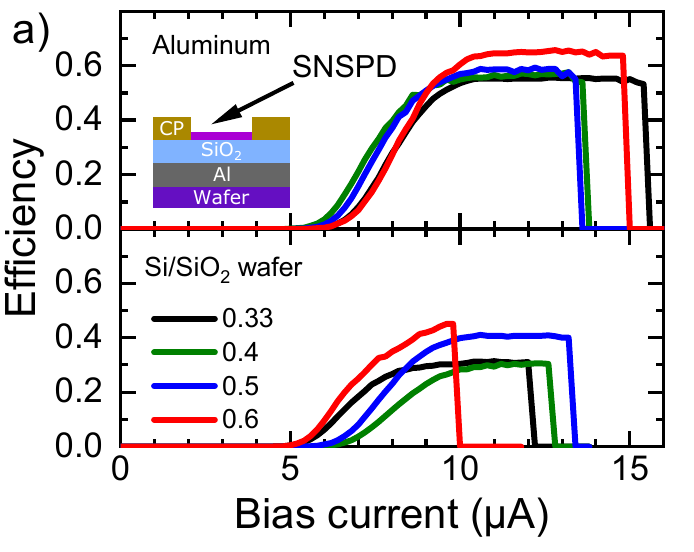}
\includegraphics[height=6cm]{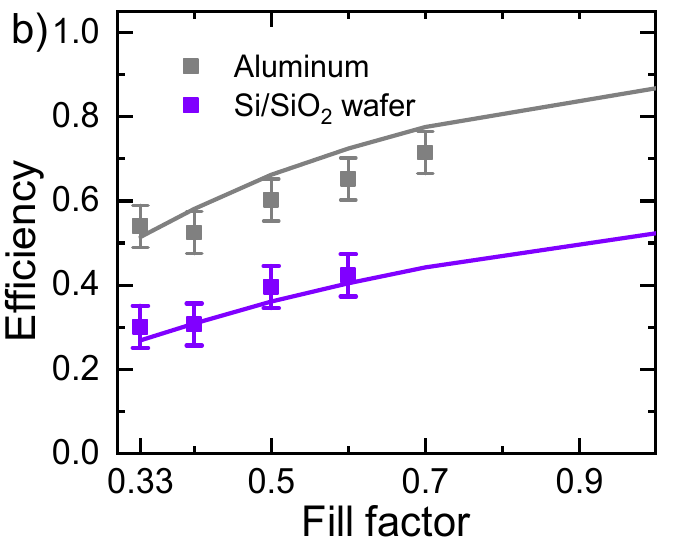}
\caption{Integration of superconducting single-photons detectors into a single-sided resonator consisting of an aluminum mirror and a SiO$_2$ spacer. (a) Detection efficiency of superconducting detectors as a function of the applied bias current at a wavelength of \SI{780}{\nano\meter}. 
The data shows detectors with fill factors 0.33 (black), 0.4 (green), 0.5 (blue), and 0.6 (red).
The bottom panel shows results for detectors fabricated directly on top of the Si/SiO$_2$ wafer, while the top panel shows the results for the aluminum semi-sided resonator.
The detection efficiency increases with increasing current until it starts to saturate from around \SI{10}{\micro A} indicating an internal quantum efficiency of one. The typical switching current ranges between \SIrange{12}{15}{\micro A} with one outlier at \SI{10}{\micro A} for a fill factor of 0.6.
While efficiencies up to \SI{40}{\percent} were achieved on top of the wafer, the bottom resonator allowed to improve the efficiency up to \SI{70}{\percent}.
The inset shows a sketch of the multi-layer structure used for the semi-sided resonator including the wafer, the aluminum mirror, the SiO$_2$ spacer, the superconducting detector (SNSPD) and the contact pads (CP) used for electrical contacting the device.
(b) Summary of the measured detection efficiencies as a function of the fill factor at a wavelength of \SI{780}{\nano\meter}. In agreement with simulations, the aluminum mirror improves the absorption and therefore detection efficiency significantly compared to a silicon dioxide-on-silicon wafer. 
The solid lines correspond to the simulation results shown before at a fixed wavelength of \SI{780}{\nano\meter}.
}
\label{fig:6}
\end{figure}

Finally, Fig. \ref{fig:6} shows the measured detection efficiencies of detectors fabricated on top of a silicon dioxide-on-silicon wafer and on top of a silicon dioxide-on-aluminum bottom resonator.
Fig. \ref{fig:6}(a) presents the measured detection efficiency for \SI{6}{\nano\meter} thick meandering NbTiN detectors with a buried aluminum mirror in the top panel and without a mirror in the bottom panel as a function of the applied bias current.
Here, the fill factor was varied between 0.33 (black) and 0.6 (red).
The devices were measured at \SI{4.5}{\kelvin} and operated at a wavelength of \SI{780}{\nano\meter}.
The inset in the top panel shows a sketch of the multi-layer structure used.
Here, the detector (SNSPD) is embedded in a semi-sided resonator structure composed of the silicon wafer, the aluminum mirror, and the SiO$_2$ spacer.
The titanium/gold contact pads (CP) are used for electrically contacting the device.
Both presented data sets show an increasing detection efficiency with an increasing bias current from \SIrange{5}{10}{\micro A}, where the measured efficiency start to saturate.
This behavior is an indication of an internal quantum efficiency of one \cite{Zol23}.
This saturation persists until the switching current is exceeded, in the range between \SIrange{12}{15}{\micro A} with one outlier at \SI{10}{\micro A} for a fill factor of 0.6.
The higher switching current of the detectors on the aluminum mirror indicate that the detector fabrication worked slightly better for this sample.
These superconducting films on top of the Si/SiO$_2$ wafer (on the Al/SiO$_2$ mirror) had a critical temperature of \SI{8.9}{\kelvin} (\SI{8.4}{\kelvin}), a sheet resistance of \SI{630}{\ohm} (\SI{590}{\ohm}) measured at \SI{20}{\kelvin} \cite{Smi58}, and a residual resistance ratio \cite{Bar16} of 0.88 (0.86).
Note that a further increase in sheet resistance and decrease in critical temperature have been found to enhance the sensitivity to single photons \cite{Zol23}, particularly for longer wavelengths \cite{Sem05}.
%Note that a higher sheet resistance and a lower critical temperature would enhance the single-photon sensitivity, especially for longer wavelengths \cite{Sem05, Zol23}.
%these parameters strongly influence for instance the cutoff wavelength \cite{Sem05} ($\lambda_\text{cutoff}  \propto \frac{R_\square}{T_\text{c}^2 w}$) describing the maximum wavelength that can be efficiently detected.
%It is proportional to the sheet resistance and anti-proportional to the critical temperature and nanowire width.
%Hence, this proportionality indicates how to adjust these parameters to improve the internal detection efficiency.
In addition, the data reveal an efficiency up to around \SI{40}{\percent} on top of the silicon dioxide-on-silicon wafer with a SiO$_2$ thickness of \SI{127}{\nano\meter}.
For the same detectors fabricated on top of the Al/SiO$_2$ semi-sided resonator, efficiencies up to \SI{70}{\percent} were measured.
Here, an optimized aluminum mirror with a thickness of \SI{50}{\nano\meter} buried underneath a \SI{120}{\nano\meter} thick SiO$_2$ layer was used.
Fig. \ref{fig:6}(b) depicts a summary of the efficiency as a function of the fill factor.
A significant improvement of the detection efficiency is observed, in FDTD simulations of the absorption (solid lines), which represent the upper limit of the theoretically achievable detection efficiency corresponding to the simulation results shown in Fig. \ref{fig:5}(b).
More detailed information about the FDTD simulations can be found in the Methods section.
To conclude, we were able to boost the efficiency from around \SI{40}{\percent} up to \SI{70}{\percent} by only using a broadband semi-sided resonator as an SNSPD substrate. 

\section*{Conclusion}
In this study, we investigated the growth conditions for ultra-thin aluminum films that can be used for various applications ranging from the aerospace industry to quantum devices.
To obtain high-quality films, we developed the sputtering process shown in Tab. \ref{tab:3}.

\begin{table}[ht]
\centering
\begin{tabular}{|l|l||l|l|}
\hline
\textbf{Parameter} & \textbf{Optimized value} & \textbf{Parameter} & \textbf{Optimized value}  \\
\hline
Temperature & \SI{20}{\celsius} & Distance & \SI{40}{\milli\meter} \\ \hline
Thickness & \SI{50}{\nano\meter} & Ar pressure & \SI{2E-3}{\milli\bar} \\ \hline
Configuration & RF & Power & \SI{200}{\watt} \\ \hline
\end{tabular}
\caption{\label{tab:3} Overview of the optimized parameter set used to sputter the high quality aluminum films.}
\end{table}

It should be noted, however, that these optimized values may differ slightly for different sputtering systems.
By tuning parameters such as the deposition rate, temperature or power, we were able to produce \SI{50}{\nano\meter} Al thin films with a roughness less than $\SI{1}{\nano\meter}$ and high reflectivity.
We also found that the roughness of the films is directly linked to the grain size and, thus, were able to improve the surface roughness by reducing the grain size.
%We also found that the film roughness is directly related to the grain size.
%Therefore, by reducing the grain size, we improved the surface roughness.
Subsequently, we added a SiO$_2$ layer and fabricated a superconducting nanowire single-photon detector on top.
This procedure allowed us to reduce the roughness down to \SI{0.25}{\nano\meter}, which is similar to the roughness of a silicon wafer.
In agreement with FDTD simulations, we were able to boost the detection efficiency at an operation wavelength of \SI{780}{\nano\meter} from \SI{40}{\percent} up to \SI{70}{\percent} by only using a single-sided resonator underneath the device.
Furthermore, note that it is possible to tune the design wavelength of the resonator by adjusting the SiO$_2$ thickness (cp. Fig. \ref{fig:5}(b)). 
For even longer wavelengths, thicker dielectric layers are required, which simultaneously allow to further reduce the overall surface roughness.

\section*{Methods}
\label{sec:Methods}
\subsection*{DC and RF magnetron sputtering}
\DeclareSIUnit[number-unit-product = {}]{\inchQ}{\textquotedbl}
\DeclareSIUnit[number-unit-product = {\thinspace}]{\inch}{inch}

For the deposition of our thin film materials, including Al, we use a BESTEC UHV Sputter-Deposition-System, containing multiple 2 inch and 3 inch sputter targets with DC and RF power supplies.
To produce contaminant-free films we pump the system down to a base pressure \SI{<1e-8}{\milli\bar}. %and the temperature controlled stage is at ambient temperature. 
Subsequently, a sputtering atmosphere of pure Ar gas and constant pressure is realized by a down-stream pressure control system. Next, the Ar plasma is ignited by either a DC or RF power supply, after which the deposition rate is determined via a thickness sensor giving full control of the film thickness.
Beyond that, the substrate is continuously rotated during the sputter process to obtain a homogeneous aluminum film.
%For the deposition of the SiO$_2$ layers, we used the following set of parameters.

%\subsection*{Sputtering routine to fabricate high-quality aluminum thin films}
%In addition to the steps described above, we developed a specific sputter recipe, which allows the deposition of a spike-free, \SI{50}{\nano\meter} thick Al film with a RMS roughness \SI{<1}{\nano\meter}.
%Prior to the deposition process, we remove any contaminants from the \SI{3}{\inchQ} target such as AlN.
%To do so, we perform a \SI{10}{\minute} cleaning process while the shutters remain closed.
%Here the sputter parameters are identical to the deposition process.
%After the stage has cooled down again and the chamber has reached a base pressure \SI{<1e-8}{\milli\bar}, we flood the chamber with argon until the desired working gas pressure is reached.
%Further, we position the substrate as close a possible to the sputter target, again leading to an increased deposition rate.
%Finally, the plasma is ignited using a RF power supply.

\subsection*{Film characterization and analysis}
To analyze the reflectivity of the deposited films, a reflectometer (Filmetrics F20) was used.
The reflectometer covers a broad spectral range  from \SIrange{200}{1000}{\nano\meter}.
The beam was focused on the sample surface.
The reflected light was then analyzed by a spectrometer, which can resolve intensities of light with a spectral accuracy below \SI{1}{\nano\meter}.
The surface roughness was measured with the help of an atomic force microscope (AFM, Asylum Research MFP-3D) operated in tapping mode and an AC240TS-R3 tip. The AFM had a resolution limit of around \SI{0.25}{\nano\meter} in roughness.
%To analyze the properties of the deposited thin films, such as reflectivity and roughness, we used a reflectometer (Filmetrics F20) as well as an atomic force microscope (AFM, Asylum Research MFP-3D).
The AFM scans were analyzed using the open-source software Gwyddion. %\cite{Necas2012}}.
Here, the RMS surface roughness is defined as $R_{rms} = (N^{-1} \sum_{i=1}^{N} (z_{i}-\bar{z})^2)^{1/2}$ with the number of scan points $N$ and their respective height $z_{i}$.
Spikes were defined by introducing a height threshold (\SI{10}{\nano\meter} above the mean level).
Lastly, the mentioned grains are defined and measured via the so-called watershed segmentation algorithm \cite{Chu12} allowing to analyze the structural properties.

\subsection*{Detector fabrication and characterization}
After the deposition of the different materials including Al, SiO$_2$ and NbTiN, we pattern the devices by electron beam lithography (EBL) using a negative-tone eBeam resist.
After development, the design is transferred into the NbTiN layer via a dry etching process in a reactive ion etching (RIE) system.
Subsequently, the geometry of the electrical contact pads is patterned using optical lithography and subsequently developed. % AZ5214E resist used in reversal mode by optical lithography using a maskless aligner.
Finally, thin layers of titanium and gold are evaporated.
Afterward, a liftoff process is used to remove excessive material.
%Finally, titanium gold contact pads are evaporated and a liftoff process is used to remove excessive gold.
The detectors were characterized in a cryogenic probe station from Janis operated at \SI{4.5}{\kelvin} sample surface temperature.
Here, a calibrated parallel beam is used that allows flood illumination and fast characterization of multiple devices \cite{Fla23a}. 
By broadening the beam to a diameter of about \SI{500}{\micro\meter} using lensed optics, a homogeneous beam spot is generated with a reduced incident photon flux per \SI{10}{\micro\meter} $\times$ \SI{10}{\micro\meter} by means of geometric attenuation.
After calibrating the ratio of the light impinging on the detector area to the overall incident photon flux by performing a 2D scan over the sample with the detector (typically \SI{5}{\milli\meter} $\times$ \SI{5}{\milli\meter}) in the center, the detection efficiency can be determined.
Thus, positioning the laser with submicron precision can be avoided as small deviations in position do not correspond to significant changes in the incident photon flux on the active detector.
Subsequently, the measured photon count rate is compared with the number of incoming photons on the detector area considering the previously mentioned geometric attenuation.
Finally, the measured optical properties are compared to finite-difference time-domain (FDTD) simulations using the commercially available software Lumerical (Ansys).
Here, the optical response of an SNSPD is approximated by a single nanowire section embedded in an optical resonator with the appropriate optical material constants and periodic boundary conditions.
This is possible since the detector geometry is symmetric in the active area of the detector, which reduces simulation times.
%\textcolor{green}{For the citations: Capitalization is turn off here, probably depends on the citation style? which one should we use here? In case of manual override is needed: https://tex.stackexchange.com/questions/10772/bibtex-loses-capitals-when-creating-bbl-file. Also: Titel is missing once, vgl. with notes from matthias}\textcolor{red}{Might try this during the submission process..}

\section*{Author contributions statement}

R.F. conceived the experiments and drafted manuscript. R.F., and C.S. conducted the experiments and analysed the results. R.F., C.S. and L.Z. fabricated samples. M.M. supported the thin film deposition processes and growth optimization. M.A. and R.G. led the research work related to thin film deposition technology. J.F. and K.M. led the research projects. All authors discussed the results and reviewed the manuscript. 

\section*{Competing interests}
The authors declare no competing interests.

\section*{Funding}
We gratefully acknowledge the German Federal Ministry of Education and Research via the funding program Photonics Research Germany (contract number 13N14846), via the funding program quantum technologies - from basic research to market (contract numbers 16K1SQ033, 13N15855, 13N15982, 13N16214 and 13N15760), and via the projects Q.com (contract number 16KIS0110) and MARQUAND (contract number BN105022), as well as the Deutsche Forschungsgemeinschaft (DFG, German Research Foundation) under Germany’s Excellence Strategy – EXC-2111 – 390814868. This research is part of the Munich Quantum Valley, which is supported by the Bavarian state government with funds from the Hightech Agenda Bayern Plus.

%%%END OF MAIN TEXT%%%


%\balance


%%%REFERENCES%%%
\bibliography{document} %You need to replace "rsc" on this line with the name of your .bib file
\bibliographystyle{spiebib} %the RSC's .bst file

\end{document}